\documentstyle[aps,prl,epsf,floats]{revtex}

\newcommand{\be}{\begin{equation}}
\newcommand{\ee}{\end{equation}}
\newcommand{\bea}{\begin{eqnarray}}
\newcommand{\eea}{\end{eqnarray}}
\newcommand{\beann}{\begin{eqnarray*}}
\newcommand{\eeann}{\end{eqnarray*}}
\newcommand{\bma}{\begin{array}{cc}}
\newcommand{\bmaccc}{\begin{array}{ccc}}
\newcommand{\ema}{\end{array}}

\newcommand{\fr}{\frac}
\newcommand{\df}{\stackrel{\rm def}{=}}

\newcommand{\la}{\langle}

\newcommand{\nn}{\nonumber}

\newcommand{\xsi}{\xi}

\newcommand{\ba}[1]{\mbox{$\begin{array}{#1}$}}
\newcommand{\ea}{\end{array}}
\newcommand{\li}{\left}
\newcommand{\re}{\right}

\begin{document}

\ifpreprintsty \else
\twocolumn[\hsize\textwidth\columnwidth\hsize\csname@twocolumnfalse%
\endcsname \fi

\draft

\title{Quantum chaos in optical systems: The annular billiard}

\author{Martina Hentschel$^1$$^*$ and Klaus Richter$^2$}
\address{
$^1$ Max-Planck-Institut f\"{u}r Physik komplexer Systeme,
N\"{o}thnitzer Str. 38, 01187 Dresden, Germany\\
$^2$ Institut f\"ur Theoretische Physik, Universit\"at Regensburg,
93040 Regensburg, Germany
}

\date{\today}
\maketitle

\begin{abstract}
We study the dielectric annular billiard as a quantum chaotic model of a
micro-optical resonator. It differs from conventional billiards with hard-wall
boundary conditions in that it is partially open and composed of two dielectric
media with different refractive indices. The interplay of reflection and
transmission at the different interfaces gives rise to rich dynamics of
classical light rays and to a variety of wave phenomena. We study the ray
propagation in terms of Poincar\'e surfaces of section and complement it with
full numerical solutions of the corresponding wave equations.
We introduce and develop an $S$-matrix approach to open optical cavities which
proves very suitable for
the identification of resonances of intermediate
width that will be most important in future applications like optical
communication devices. We show that the Husimi representation is a useful tool in
characterizing resonances and establish the ray-wave correspondence in
real and phase space. While the simple ray picture provides a good
qualitative description of certain system classes, only the wave description reveals
the quantitative details.

\end{abstract}

\ifpreprintsty \else
] \fi              


\section{Introduction}
Billiard systems of many kinds
have proven to be very fruitful model systems in the field
of quantum chaos. The methods of investigation are well established both for the
classical dynamics and for the quantum mechanical behaviour, with
semiclassical methods describing the transition from quantum to classical
properties. With the growing interest in quantum chaos and in mesoscopic
physics, new systems have entered the stage, including systems exhibiting chaos
of classical waves such as (macroscopic) microwave billiards
\cite{stoeckmann,achimrichter},
acoustic resonators \cite{acousticreso} as well as deformed microcavities
\cite{noeckelstone,optlettschnucki,wilkinson,sangwook,doya,troepfchen}
which can operate as microlasers \cite{backes,gmachl}.
To describe these (two-dimensional) systems one can exploit the analogy
between the stationary Schr\"odinger
equation and the Helmholtz equation for (classical) waves \cite{jackson}.
Quantum chaotic experiments using
microwave cavities or other classical waves
(e.g., acoustic or water waves) are based on this
mathematical equivalence, see \cite{stoeckmannbuch} for a review.
Most of the investigated systems are hard-wall billiards.
However, for the class of optical, or dielectric,
model systems 
the billiard boundary manifests
itself by a change in the index of refraction allowing for reflection and transmission
of light. The limit of closed systems is
approached as the difference in the refractive indices reaches infinity.

We emphasize that the openness of optical systems extents the set of interesting
questions with respect to those for closed billiards.
In this paper we suggest a further extension of the class of open optical cavities
by considering two regions with different refractive indices inside the cavity,
which leads to an additional refractive interface between the two dielectrics
inside the resonator. The interplay between refraction inside the billiard and
partial reflection at the outer billiard boundary gives rise to a variety of
phenomena 
in the classical ray dynamics and correspondingly in the wave description of
such systems.

The model we study is the annular
billiard shown in Fig.~\ref{fig_annbgeom}.
It consists of a small disk of radius $R_2$
placed 
inside a larger disk of radius $R_1$ with a 
displacement $\delta$ of the disk centres. This system is well-known from quantum
and wave mechanical studies of the
hard-wall configuration \cite{bohigas,doronfrischat,dembowski}
with non-vanishing
wave-functions only in the annular region. 
It carries features
of a ray-splitting system \cite{bluemel} when each disk is characterized by a
(stepwise constant) potential (unlike the situation we will discuss here, see
Sec.~\ref{subsec_maxwschroed}). 
Here we consider disks characterized by indices of refraction, $n_1$ and
$n_2$, respectively, with the index of the
environment fixed at $n_0=1$ \cite{remark_doublet}.
We will study 
billiard materials with $n_i>1$ such that confinement by total internal reflection
is possible. Then methods well-known from the description 
of classical dynamical systems, such as the use of  
Poincar\'e's surfaces of section, 
can be employed to describe the ray
dynamics. Note that whispering gallery modes in the dielectric annular billiard
with a metallic inner disk have been discussed in
Ref.~\onlinecite{hackenbroich}. A detailed study of periodic orbits in a
specific hard-wall configuration, together with the expected consequences
on the electromagnetic scattering problem was performed in
Ref.~\onlinecite{gouesbet1}. 

\begin{figure}[!t]
\epsfxsize=8.5cm
\centerline{\epsffile{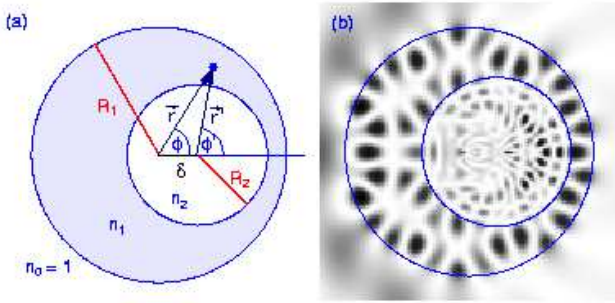}}
  \begin{center}
  \caption{(a) Geometry and notations of the dielectric annular billiard.
  (b) Example of a resonant wave function excited by a plane wave with wave number
  $k = 6.251$ incident
  from the left
  ($R_1=1, R_2=0.6, \delta=0.22, n_1=3, n_2=6$). Dark regions denote high
  electric field intensity.
  }
  \label{fig_annbgeom}
   \end{center}
\end{figure}

The above-mentioned correspondence between the Helmholtz and Schr\"odinger
equation is established by means of an effective potential \cite{effpot}
that depends not only on
position and the respective index of refraction $n$, but also on the energy.
Changes in the refractive index 
give rise to steps in the 
effective potential which allows for a
quantum-mechanical interpretation (e.g., quasibound states, tunneling escape).
We will discuss this point in the next section when we contrast optical
systems governed by Maxwell's equation with quantum mechanical problems obeying
the Schr\"odinger equation. 
Also, we will see
how the two possible polarization directions affect the Maxwell-Schr\"odinger
correspondence and which quantity takes the role of $\hbar$: Maxwell's
equations are, of course, not aware of the existence of $\hbar$.

The further outline of the paper is as follows:
In Sec.~\ref{sec_disk}
we introduce the ray and wave optics notion 
for the simple
system of the dielectric disk that arises from the annular system upon
removal of the inner obstacle ($n_1=n_2$, or $R_2=0$).
We describe the methods used for the study
of the annular billiard in the subsequent sections; 
namely the adaptation of the  
Poincar\'e surface of section method, well-known from classical mechanics, to optical
systems, the exact solution of the Maxwell 
equation (leading to an effective Schr\"odinger equation),
and the $S$-matrix approach.  
Whereas the first approach is based on the ray picture,
the latter two clearly fully include the wave nature
of light.
We employ 
these methods for the annular billiard in
Sec.~\ref{sec_annbpictures}, 
where we introduce methods to study the ray dynamics in optical compound
systems
and apply, for the first time
to our knowledge, an $S$-matrix formalism to 
optical billiards.
The expected ray-wave, or classical-quantum, correspondence is established in
Sec.~\ref{sec_annbraywave} and investigated 
from various viewpoints,
including both real space and phase space
arguments. 
However, several features in the behaviour of waves require improvements
of the simple ray model as we will illustrate and explain with typical examples.
In our conclusion, Sec.~\ref{sec_concl}, 
we discuss the possibility of an experimental realization of the annular
system with the currently available dielectric materials. 
The successes of the ray-picture illuminated here and
elsewhere \cite{noeckelstone,optlettschnucki,troepfchen,gmachl,optlett}
suggest the ray-based design of micro-optical cavities for, e.g., future 
communication technologies.


\section{The dielectric disk}
\label{sec_disk}

In this section we introduce the methods, techniques, and notations
used later in the discussion of the annular billiard. We present 
the ray and wave picture for the description of optical (or dielectric)
systems using the simple example of a dielectric disk,
which provides all the ingredients 
to deal with the
annular billiard (apart from a coordinate transformation,
see below). We start with the ray optics approach and
show how methods well established in classical dynamics can be adopted to
optical systems. In 
the wave description we distinguish
between an approach to the resonant states of the  
(naturally) open optical system by complex wave
vectors based on Maxwell's equations on the one hand, and by real wave vectors
arising in an $S$-matrix approach on the other hand.

\subsection{Ray Optics: Classical billiards with total internal reflection}

\begin{figure}[!t]
  \epsfxsize=8cm
  \centerline{\epsffile{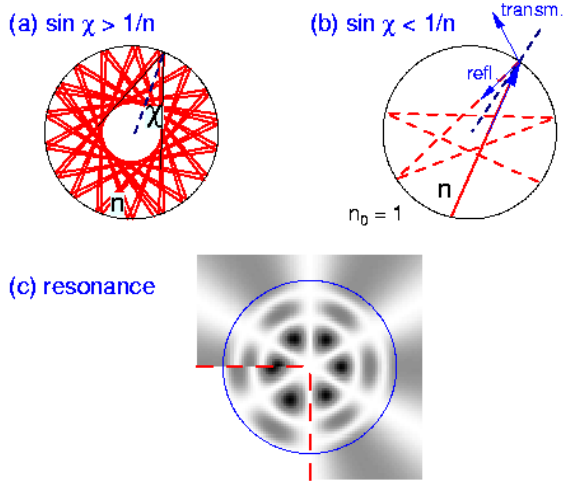}}
  \begin{center}
  \caption{
  Comparison of ray and wave picture for a dielectric disk of refractive
  index $n$. The upper panels illustrate the two possibilities
  of (a) total internal reflection, $\sin \chi \ge \sin \chi_c
  =1/n$, when the ray dynamics is equal to that of a classical point particle in a closed
  system, and (b) ray refraction when,
  due to partial transmission, the light intensity inside the disk decreases
  with time.
  In (c) 
  the intensity of the electric field
  (see Sec.~\ref{subsec_maxwschroed}, 
  higher intensity in darker regions) for a quasibound
  state ($nkR = 11.428 - 0.254 {\rm i}$) of the dielectric disk ($n=3$) is shown.
  For comparison, an eigenstate
  of the closed disk ($nkR = 9.761$, vanishing intensity outside the disk)
  is given in the lower left sector (both are for transverse magnetic field,
  TM polarization).
  }
  \label{fig_raywavedisk}
  \end{center}
\end{figure}

Within ray optics, the zero-wavelength limit of wave optics, light is described
by a ray that follows a straight line through a medium,
very similar to the dynamics of a point mass.
Let us assume a light ray, or plane wave, incident under an angle $\chi_1$
with respect to the normal of
a dielectric boundary where the refractive index changes from $n_1$
to $n_2$.
At the interface, the ray is i) specularly reflected under an angle
$\chi_2 = \chi_1 \equiv \chi $, with a polarization-dependent \cite{remark_polar}
probability
$R^{\rm TM /TE }$, see Fig.~\ref{fig_raywavedisk}(b).
The remaining part, $T^{\rm TM / TE} =1 - R^{\rm TM / TE}$, is ii) transmitted into
the other medium under an output angle $\eta$ given by Snell's law,
$\sin \eta = (n_1/n_2) \sin\chi \equiv n \sin \chi$. In the last identity we have employed the
scaling properties of the system that allow one to fix one of the refractive indices,
(e.g. that of the environment) to unity without loss of generality.

\begin{figure}[!t]
  \epsfxsize=8cm
  \centerline{\epsffile{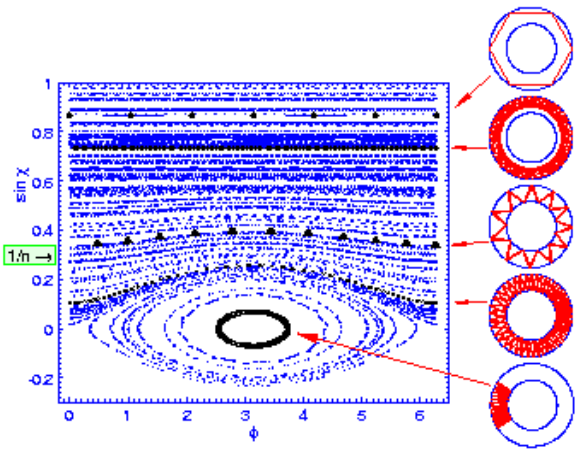}}
  \begin{center}
  \caption{
  Poincar\'e surface of section for the annular billiard with $R_1 = 1, R_2 = 0.6$, and
  $\delta=0.01$. 
  The horizontal axis 
  is the polar angle $\phi$, the vertical axis, 
  $\sin \chi$, is proportional to the angular momentum in
  $z$-direction (perpendicular to the system plane). 
  Although the displacement of the inner disk is rather
  small, it has major impact on trajectories that explore the region
  $\sin \chi < R_2 + \delta$. Other trajectories are not influenced and
  identical to those of a single dielectric disk, 
  see e.g. the two upper-most examples on the right.
  The initial angular momentum of the trajectories shown is positiv
  ($\sin \chi > 0$). 
  Nonetheless regions where $\sin \chi < 0$ are explored, implying a {\em change
  in the sense of rotation} as, e.g., in the lower-most trajectory on the
  right. The critical angle for total internal reflection is marked on the left
  indicating that the two lower-most trajectories are not confined.}
  \label{fig_sosannb01}
  \end{center}
\end{figure}

Snell's law cannot be fulfilled to yield real
$\eta$ for any angle of incidence $\chi$ if $n>1$ ($n_1 > n_2$, respectively).
{\em Total internal reflection} occurs if $\sin \chi \ge \sin \chi_c \equiv 1/n$
where we introduced the critical angle $\chi_c$. For angles of
incidence above the critical angle, light is confined by total internal reflection with
zero transmission and behaves like a classical point particle, Fig.~\ref{fig_raywavedisk}(a).
Therefore, real and phase space methods
from classical mechanics (such as ray tracing or
the Poincar\'e surface of section technique),
prove to be very useful if they are complemented by the
optical property of refraction:
The Poincar\'e surface of section (SOS) method
works exact (except for the exponentially small tunneling losses)
as long as we are
in the regime of total internal reflection. However, for $-1/n < \sin \chi
<1/n$, light can escape so that the intensity 
remaining inside
the disk is diminishing. This fact has to be taken into account when discussing
the Poincar\'e SOS 
for optical systems. Figure \ref{fig_sosannb01} shows an example of a
Poincar\'e SOS for a hard-wall annular system with slightly eccentric inner disk
($\delta=0.01$). The critical value $\sin \chi = 1/n$ is marked by an
arrow.

Probing the phase space structure of a rotational invariant system like a disk
in terms of a Poincar\'e SOS gives a uniform structure as shown in the
upper part of Fig.~\ref{fig_sosannb01}. Although this is a Poincar\'e SOS for
an annular system \cite{gouesbet2} 
with a slightly eccentric inner disk (see Sec.~\ref{sec_annbraywave}),
it is identical to that for a disk for trajectories that do not hit the inner
disk, i.e. $\sin \chi > R_2 + \delta$ ($R_1=1$). The straight horizontal lines
directly express the conservation of angular momentum, 
that is, conservation of $\sin \chi$, and the corresponding trajectories are
referred to as whispering gallery (WG) orbits.

The reflection and transmission
probabilities, $R^{\rm TM / TE}$ and $T^{\rm TM / TE}$, are provided by
Fresnel's laws \cite{lipson}. 
A plane electromagnetic wave incident on a planar dielectric interface
with angle of incidence $\chi$ is reflected
with the polarization-dependent probabilities
\be
R^{\rm TM} =  \fr{\sin^2(\chi-\eta)}{\sin^2(\chi+\eta)}\:,\quad
R^{\rm TE}  =  \fr{\tan^2(\chi-\eta)}{\tan^2(\chi+\eta)}
\label{frestmte} \:,
\ee
where
TM (TE) denotes transverse polarization of the magnetic (electric)
field at the interface, and
$\eta = \arcsin (n \sin \chi)$ is the direction of the refracted beam
according to Snell's law.

\subsection{Wave Picture: From Maxwell to Schr\"odinger}
\label{subsec_maxwschroed}

Before we turn to the more complicated annular billiard in
Sec.~\ref{sec_annbpictures}, we
assume an infinite dielectric cylinder of
radius $R$ and refractive index $n$ embedded in vacuum 
with refractive index $n_0 = 1$.
We will call $k$ the wave number outside the cylinder
and, analogously, $ n k $ is the wave number
inside.
The solution of Maxwell's equations for the vortices of the
electromagnetic field \cite{lipson}
is given, e.g., in
Refs.~\onlinecite{noeckelthesis,effpot}, and  
leads to an equation for the electric (magnetic) field
that is {\it similar} to the conventional Schr\"odinger
equation.
The vector character of the fields implies, however,
that one has to distinguish two possible
polarization directions with differing boundary conditions.
The situation where the electric (magnetic) field is
parallel to the cylinder ($z$-) axis
is called TM (TE) polarization, with the magnetic (electric) field being thus
transverse.
Using the rotational invariance of the system, separation in cylindrical
variables (assuming a $\phi$-dependence $e^{im\phi}$,
and a $z$-dependence $e^{i k_z z}$)
eventually leads to an effective
Schr\"odinger equation \cite{remark_radschroed} for the radial component of the
electric field,
\be
\label{radschroed}
-\left[ \fr{d^2}{dr^2} + \fr{1}{r}\fr{d}{dr}\right] E(r) + V_{\rm eff}(r) E(r)
= k^2 E(r),
\ee
where we introduced the effective potential
\be\label{veff}
V_{\rm eff} (r) = k^2 (1-n^2) + m^2/r^2 + k_z^2 \:.
\ee
The first term reveals
immediately that dielectric regions with $n > 1$ correspond
to an attractive well in the quantum
analogy, and that a potential structure is determined by the 
change of the refractive indices for different regions.
Note, however, the {\it energy-dependent} prefactor -- a far-reaching difference
in comparison to quantum mechanics.
The other two terms in Eq.~(\ref{veff}) arise from the conservation of the
angular momentum
(characterized by the quantum number $m$), and express the
conservation of the linear momentum along the cylinder axis (acting
as an offset in energy), respectively.

In the following we will consider a dielectric disk (that is, we choose a
particular cross sectional plane of the cylinder to obtain
an effective system), and set $k_z$ to zero
corresponding to a wave in the $x$-$y$ plane.
Approaching the disk 
from the outside ($r>R, n_0 = 1$) there is
only the angular momentum contribution 
to the effective potential $V_{\rm eff}$, Eq.~(\ref{veff}).
At $r=R$, there is a discontinuity in 
$V_{\rm eff}$ that is  
proportional to $1-n^2$,
reflecting the non-continuous change in the refractive
index.
It reaches from $k_{\rm max}^2 = m^2 /R^2$ to
$k_{\rm min}^2 = m^2 /(n\, R)^2$.
Inside the disk the angular momentum contribution,
now shifted by $k_{\rm max}^2 - k_{\rm min}^2$, again
determines the behaviour (see also Fig.~\ref{fig_twowgms}
at values $r/R_1 > R_2/R_1$ ).

The form of the potential suggests an interpretation in the spirit
of quantum mechanics with metastable 
states in the potential well
that decay by tunneling escape, 
and indeed this turns out to be the quantum-mechanical version of confinement
by total internal reflection \cite{noeckelthesis,effpot}.
To this end we employ
a semiclassical quantization condition for the $z$-component of the
quantum mechanical and classical angular momentum,
$m \hbar = n \hbar {\rm Re} (k R) \sin \chi$.
We find
\be\label{sinchi}
\sin \chi = \frac {m}{n {\rm Re} (k R)} \:
\ee
as relation between the angle of incidence as ray picture quantity, and the
wave number and angular momentum of a resonance.

Another correspondence between ray and wave quantities exists between the
(polarization-dependent) Fresnel reflection coefficient $R^{\rm TM/TE}$
and the imaginary part of the wave number
that describes the decay 
of a resonant state. In fact one can deduce
a reflection coefficient $R_d^{\rm TM/TE}$ of the disk \cite{procamst},
\be
R^{\rm TM / TE}_d  = \exp [4 n \,{\rm Im} (k R) \cos \chi ]\:.
\label{rcoeff}
\ee
We wish to point out that there exist deviations between
the Fresnel values $R^{\rm TM/TE}$
and $R^{\rm TM/TE}_d$ when the wavelength becomes comparable to the system
size, in particular around the
critical angle. This can be understood
within a semiclassical picture based on the Goos-H\"anchen
effect \cite{fresnelpre,ghslit}.

The general solutions of the radial Schr\"odinger equation (\ref{radschroed}) \
are Bessel and Neumann functions, 
$J_m(k_i r)$ and $Y_m(k_i r)$ of order $m$, where
$k_i$ is the wave number in the respective medium.
Since physics
requires a finite value of the wave function at the disk center, the solution
inside the disk can consist of Bessel functions only. Outside the dielectric we
assume an outgoing wave function, namely a Hankel function $H_m^{(1)}$ of the
first kind, in accordance with our picture of
a decaying state.
The resonant states are obtained by matching the
wave field $\propto J_m(nkr)  e^{i m \phi}$ inside the disk
at $r=R$
to the wave field $\propto H_m^{(1)}(kr) e^{i m \phi}$ outside the disk
according to the  polarization
dependent matching conditions deduced from Maxwell's equations.
The resonant states are solutions of
\be \label{tmstartderiv}
J_m(nkR) \, {H_m^{(1)}}'(kR) = {\cal P} \, J'_m (nkR) \, H_m^{(1)}(kR) \:,
\ee
where ${\cal P} = n$ ($1/n$) for TM (TE) polarization
(primes denote derivatives
with respect to the full arguments $nkr$ and $kr$, respectively).

One example of a quasibound state as solution of the optically open system
is shown in Fig.~\ref{fig_raywavedisk}(c) and
compared to a solution of the closed disk. The shift in the wave patterns is
clearly visible.
Owing to the symmetry of the system we find a characteristic
(quantum-mechanical) node structure that is directly related to the quantum
numbers $m$ (there are 2$m$ azimuthal nodal points), and $\rho$ counting the
radial nodes (hence $m=3, \rho=2$ in the example).

At this point a further discussion concerning the
appearance of $\hbar$ in optical systems is convenient.
Employing the quantum-classical correspondence, one
expects $\hbar$ to be related to the reciprocal wave number,
$\hbar \sim 1 / k $, because $\hbar \to 0$ in the classical (here the ray)
limit $k \to \infty$. 
This relation is indeed obtained when we compare
Eq.~(\ref{radschroed}), divided by $k^2$ (thereby removing the energy-dependence
of the effective potential), with
Schr\"odinger's equation, and identify $1/k$ with $\hbar$.

\subsection{$S$-Matrix approach to the dielectric disk}
\label{subsec_smatdisk}

The main idea when considering a scattering problem is to probe the
response of the system to incoming (test) waves, and to
extract system properties like resonance positions and widths
from  the {\em scattered} wave.
Physically, this method is formulated for
real wave vectors.

Here we want to investigate 
the scattering properties of the dielectric disk for electromagnetic waves
in the framework of 
$S$-matrix theory \cite{uzy_iod,taylor,uzyleshouches}.
One possible choice for the incident test waves are, of course,
plane waves. For
our rotational invariant disk of finite dimension,
however, incident waves that
allow for angular momentum classification are much more
convenient: Then we need to take into consideration only waves
with impact parameter of the order of the system dimension or
smaller. The Hankel functions $H_m^{(2)}$ of the second kind
possess the desired properties.

Again, we consider a dielectric disk of radius $R$ and refractive
index $n$ and denote the vacuum wave number by $k$. 
According to Maxwell's equations and the discussion in the previous
Sec.~\ref{subsec_maxwschroed} we write the
wave function $\Psi_m^{\rm scatt}$ outside
that is excited by an incident wave of angular momentum $m$
as
\[
\Psi_m^{\rm scatt} (kr)
  = H_m^{(2)} (k r) e^{i m \phi}+
                        \sum_{l=-\infty}^{\infty} S_{ml} H_l^{(1)} (k r)
                         e^{i l \phi}
                        \: .
\]
Here, $S_{ml}$ is the amplitude for 
an incident wave $H_m^{(2)}$ 
to be scattered into $H_l^{(1)}$. The scattering
amplitudes are comprised in the $S$-matrix.
It follows from flux 
conservation that $S$ has to be unitary, a property that we will
use subsequently. Starting with a general situation in which $S$
can have entries everywhere, symmetry requirements will reduce the
number of independent matrix elements.
For the dielectric
disk, the scattered wave 
has to obey angular momentum conservation and will, therefore, be
a Hankel function of the same order $m$ as the incoming Hankel
function. Hence the 
scattering matrix is diagonal. 
In the general case where angular momentum is not conserved (as, for
example, for a deformed disk or the annular geometry  that we will consider
in all following sections),
scattering will occur
into {\em all} possible angular momenta $l$. 

Employing the matching conditions
(cf.~Sec.~\ref{subsec_maxwschroed}) for TM polarization,
we obtain from
the requirement of continuity of the wave function (or the
electric field) and their derivative
eventually the matrix elements $S_{m m'}$:
\be
S_{m m'}  =  - \fr{{H_m^{(2)}}'(kr) -
                        n \fr{J'_m(nkr)}{J_m(nkr)}{H_m^{(2)}}(kr)}
                 {{H_m^{(1)}}'(kr) -
                        n \fr{J'_m(nkr)}{J_m(nkr)}{H_m^{(1)}}(kr)}
                        \: \delta_{m m'}
                 \label{scattmatdisk}   \:.
\ee

The general idea for identifying resonances 
is that a probing wave
with resonance energy 
will interact {\em longer} with the system
than a wave with ``non-fitting'' energy. 
This can be quantified in terms of
the {\em Wigner delay time} $\tau^W (E_{k})$ \cite{wignerdelay}
that is the derivative of the total phase $\theta$ of the determinant of
the $S$-matrix, det$S$=$e^{i\theta}$,
with respect to energy $E_{k}=k^2$,
\be\label{defwigdelay}
\tau^W (E_{k})= \fr{d \theta (E_{k})} {d E_{k}} \:.
\ee

\begin{figure}[!t]
  \epsfxsize=9cm
  \centerline{\epsffile{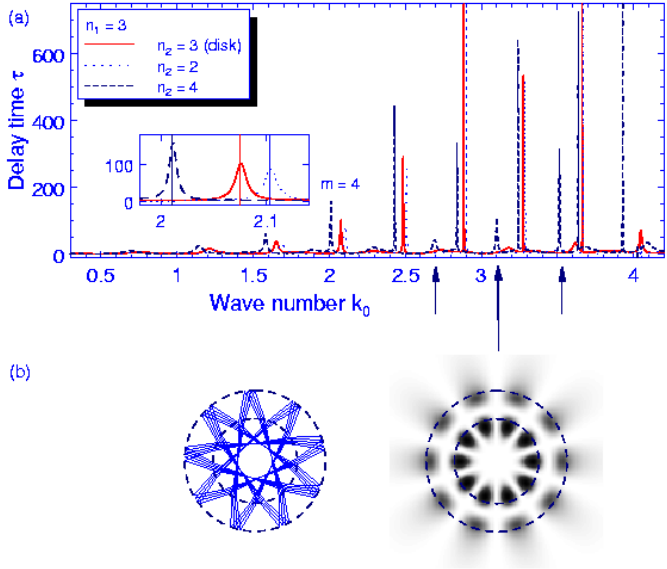}}
  \begin{center}
  \caption{(a) Resonances in the concentric annular billiard
  ($R_1=1, R_2 = 0.6, n_1 = 3$), 
  corresponding to the
  first family of whispering
  gallery modes. The annular systems $n_2 = 2$
  (dotted) and $n_2=4$ (dashed) are compared with the homogeneous disk $n_2 = 3$.
  Note the systematic deviation of the resonance position
  to larger (smaller) wave numbers for $n_2 = 2$ ($n_2 = 4$) that decreases with
  increasing angular momentum quantum number $m$ since the inner disk becomes
  less important. In the inset, we compare the 
  positions and widths of the 4th resonance ($m=4$)
  in the delay time with the respective complex wave numbers according to
  Eq.~(\ref{annbtmexact}).
  We find excellent agreement with the numerically exact values
  $k_0 = 2.0108 - 0.0041 {\rm i}$ ($n_2 = 4$),
  $k_0 = 2.0753 - 0.0063 {\rm i}$ ($n_2 = 3$), and
  $k_0 = 2.1035 - 0.0075 {\rm i}$ ($n_2 = 2$).
  Note the existence of additional resonances for $n_2 = 4$, some of them
  marked by arrows from below and illustrated in the ray and wave picture in (b).
  They are due to the double-well
  structure of the effective potential and referred to as ``double WGMs'', see text.}
  \label{fig_delayconc}
  \end{center}
\end{figure}

In the following, we will use the wave-number based delay time
\be\label{defdelaytau}
\tau (k) 
\equiv 4 \pi \, k \: \tau^W
(k^2)
\ee
in order to identify resonances as depicted in Fig.~\ref{fig_delayconc}(a).
The solid line
shows the result for a dielectric disk with $n$=3. Families of whispering gallery
modes (WGMs) are identified
upon increasing the wave number and can be labelled by the quantum number $m$
that counts the azimuthal nodes (2$m$). The decrease in peak width, accompanied by an
increase in height, that is observed with increasing $m$ corresponds to an
increase of the angle of incidence, Eq.~(\ref{sinchi}), and improved confinement
by total internal reflection.

Note the relation between the (total) phase $\theta(E_k)$
of the $S$-matrix and
the so-called resonance counting function:
$N(E_k) = \theta(E_k) / 2 \pi$, cf.~\cite{uzy_iod}.
The idea is that a resonance is encountered whenever
the phase $\theta$ of $\det S$
increases by $2 \pi$
upon increasing the energy $E_k$. 

In the following we will use the function $\tau(k)$ to determine
the resonances. Isolated resonances appear as (Lorentzian) peaks in $\tau(k)$, see
Fig.~\ref{fig_delayconc}(a), above a small background.
Information about the
imaginary part of the resonance 
is now encoded in the height and width of the
Lorentzian resonance peaks \cite{taylor}. 
We point out that the resolution of
very broad and extremely narrow resonances might be difficult, because they are
either included in the background or not captured using a finite numerical grid
interval.
However, resonances with a wide range of widths are easily identified, in particular
all resonances that are of interest for microlaser applications
are found within the $S$-matrix approach~\cite{remark_resolaser}.

The area under the curve
$\tau(k)$ is proportional to the
number of states with wave numbers smaller
than $k$ \cite{uzy_iod,uzyleshouches}.
In the case of {\em stepwise} potentials such as realized in ray-splitting billiards,
simple Weyl formulas for the smooth part of the density of states
were derived for a number of geometries \cite{bluemel}.
The application of these results to optical systems where ray splitting
is realized by refraction and transmission at refractive index boundaries
is tempting.
However, here we work with an
{\em energy dependent} effective potential, in contrast to 
the situation studied in \cite{bluemel} where only a (stepwise)
spatial dependence of the potential was assumed. Consequently,
a generalization of the formulas derived in \cite{bluemel} would
be required if one is interested in an analytical expression for
the smooth part of the density of states, which is, however, not the subject
of this work.

\section{ Annular billiard in the ray and wave picture}
\label{sec_annbpictures}

In this section we adopt the ray and wave methods explained above
to the general case of the dielectric annular billiard.
We will denote the three different regions, namely the environment (refractive
index $n_0$), the annular region ($n_1$), and the inner disk ($n_2$) by the
indices 0, 1, and 2, respectively. The corresponding wave numbers are
$k_0, k_1$, and $k_2$. Due to the scaling properties we 
fix $n_0 \equiv 1$ and one of the geometry parameters $R_1, R_2, \delta$;
we choose $R_1 \equiv 1$.
Given a set of parameters ($n_0 \equiv 1, n_1, n_2$), the same results
hold 
for the scaled set ($\widetilde{n}_0, \widetilde{n}_1=n_1 \widetilde{n}_0,
\widetilde{n}_2=n_2 \widetilde{n}_0$) for wave numbers
$\widetilde{k} = k /\widetilde{n}_0$, if the geometry is not changed.
In turn, fixing the
dielectric constants, the parameter sets ($R_1 \equiv 1, R_2, \delta$) and
($\widetilde{R}_1, \widetilde{R}_2 = R_2 \widetilde{R}_1,
\widetilde{\delta} = \delta \widetilde{R}_1$) are equivalent when
$k \rightarrow \widetilde{k} = k /\widetilde{R}_1$.

\subsection{Ray optics and refractive billiard}
\label{subsec_annbray}

\begin{figure}[!t]
  \epsfxsize=9cm
  \centerline{\epsffile{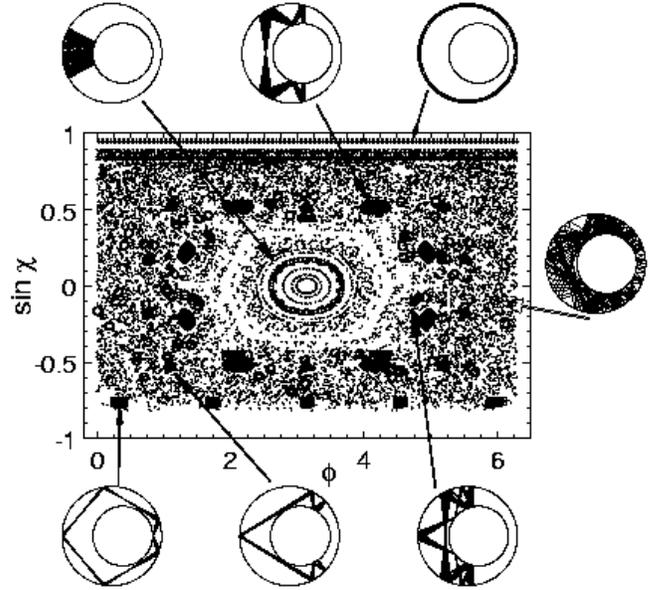}}
  \begin{center}
  \caption{Poincar\'e SOS taken at the outer boundary for the hard-wall
  annular billiard ($R_2=0.6, \delta=0.22$). Trajectories are bounded
  to the annular region, no optical properties of the system are yet included.
  Typical regular and chaotic trajectories are shown both in real
  and phase space.} 
  \label{fig_sosannbhw}
  \end{center}
\end{figure}

The rotational invariance of the circular billiard discussed
in Sec.~\ref{sec_disk} can be broken
either by deformation (as in \cite{gmachl}) or by placing off-centered
opaque obstacles inside the disk, leading to the
hard-wall annular billiard. Starting from the concentric situation,
the system stays close to integrable due to the
existence of adiabatic invariants for not too big eccentricities,
see Fig.~\ref{fig_sosannb01}. However, in general the phase space of the
annular billiard is mixed, with regular islands placed in the chaotic sea as
shown in Fig.~\ref{fig_sosannbhw}.

For optical systems, both the outer and inner boundary
become permeable. Leakage at the outer boundary occurs
for $-1/n < \sin \chi < 1/n$. In the simplest
qualitative picture, starting with the hard-wall system, we will assume 
those rays to leave the cavity (thus
simplifying Fresnel's laws (\ref{frestmte}) 
to a stepwise function). The corresponding trajectories are assumed to not
exist in an optical cavity.
If one is interested in
how the intensity of a certain trajectory decreases, Fresnel's laws can easily
be taken into account accurately, leading to the model of a Fresnel billiard
\cite{noeckelstone,optlett,berry}.

However, the description of the refractively opened
{\em inner} boundary turns out to be rather complicated.
There, all rays remain in the billiard, causing a tremendous
increase of the number of rays upon partial reflection. 
Another crucial
difference is that now new trajectories arise, namely those crossing the inner
disk. We model 
this situation by introducing the model of the ``refractive
billiard'': Whenever total internal reflection is violated at the inner
boundary,
the ray enters the inner disk according to Snell's law
with full intensity, such that now ray splitting
occurs. Otherwise, the ray is specularly reflected and stays in the annulus. This
corresponds again to a stepwise simplification of Fresnel's laws.
Note that 
the hard-wall billiard is in fact a realization of constant reflection
coefficient
$R^{\rm TM/TE}=1$. The real situation is found in between the stepwise and constant
approximations and, depending on the refractive indices chosen, results from both
limits are needed in order to understand the resonant modes found in the wave picture,
see Sec.~\ref{sec_annbraywave}.

\begin{figure}[!t]
  \epsfxsize=9cm
  \centerline{\epsffile{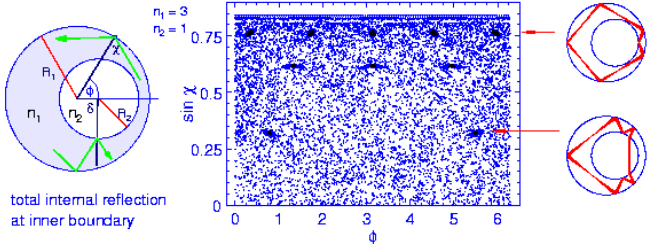}}
  \begin{center}
  \caption{
  Poincar\'e SOS taken at the outer boundary
  for the refractive billiard with $n_0=1, n_1=3, n_2=1$ and
  the same geometry as in Fig.~\ref{fig_sosannbhw}. The hard-wall condition of
  Fig.~\ref{fig_sosannbhw} is kept
  at the outer boundary, but replaced at the inner one by the condition of
  total internal reflection. If it is not fulfilled the ray will penetrate the inner disk,
  giving rise to a restructured phase space and new regular orbits as the 
  one on the lower right.} 
  \label{fig_refr131}
  \end{center}
\end{figure}

\begin{figure}[!t]
  \epsfxsize=9cm
  \centerline{\epsffile{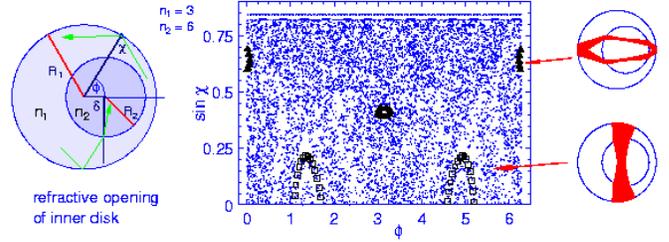}}
  \begin{center}
  \caption{
  Poincar\'e SOS for the refractive billiard with $n_0=1, n_1=3, n_2=6$ and
  the same geometry as in Fig.~\ref{fig_sosannbhw}.
  Rays in the annulus that hit the inner boundary
  will penetrate it. Note the existence of whispering gallery modes in the inner disk
  (confined by total internal reflection) not visible in this SOS.}
  \label{fig_refr136}
  \end{center}
\end{figure}

We complete our refractive-billiard model by first assuming specular
reflection at the outer boundary, and discuss outer-boundary losses
subsequently as outlined above. Results are shown in Figs. \ref{fig_refr131} and
\ref{fig_refr136} for the same geometry as in 
Fig.~\ref{fig_sosannbhw} \cite{remark_geomparam},
and two different combinations of refractive indices. In Fig.
\ref{fig_refr131}, the annular index $n_1$ is highest, allowing for total internal
reflection at both boundaries. In the limit $n_1 \to \infty$ we would recover
the phase space of the hard-wall billiard, Fig.~\ref{fig_sosannbhw}. For
moderate $n_1=3$ ($n_0=n_2=1$) as in Fig.~\ref{fig_refr131} we are,
however, away from this limit:
most of the regular trajectories of the hard-wall system are gone and,
in turn,
new regular orbits passing through the inner disk appear. 

The situation changes once more for
$n_0 < n_1 < n_2$,
because then total internal reflection at the
inner boundary is {\em never} possible (again, we base our discussion on rays
entering from the annulus), and all rays hitting the inner boundary will enter.
Furthermore, they will leave the inner disk upon the next reflection according
to the principle of reversibility of the light path. Note, however, that
confinement by total internal reflection in the {\em inner}
disk is well possible. From our discussion in Sec.~\ref{sec_disk} we know that
these orbits will and can only be whispering gallery modes (WGMs).
To anticipate results of the next section, those modes do exist and leave their
signature as very sharp peaks in the delay time.

In Fig.~\ref{fig_refr136} an example of the phase space is given, showing
yet another structure owing to the change in the refractive indices. For
the regular orbits shown at the right, we expect only the upper one to survive
the (optical) opening of the outer boundary as long as $n_0/n_1 \geq 3.2$.
The lower orbit hits the outer boundary perpendicular ($\chi=0$)
at least in some points, and
can therefore only be confined by hard walls.

\subsection{Wave picture: Maxwell's equations and $S$-matrix approach}
\label{subsec_annbwave}

Generalizing the wave picture approaches presented in Sec.~\ref{sec_disk}
for the dielectric disk to the annular billiard requires essentially
to consider another, off-centered circular boundary at which the matching
conditions resulting from Maxwell's equations have to be fulfilled as well.
An eccentric inclusion lowers the
rotational symmetry of the system to axial reflection invariance about the
symmetry axis of the system. Consequently, angular momentum is not conserved,
and the $S$-matrix of the compound system cannot be diagonal in the general
case.

Maxwell's equations can be solved analytically in the concentric case ($R_2 >0$),
and  resonant states with complex wave number are obtained as zeros of
the expression
\bea
  \label{annbtmexact}
& & n_1 J_m (k_2 R_2) {H_m^{(1)}}' (k_0 R_1) \times \\
& & \times \li[
{H_m^{(2)}}' (k_1 R_2) H_m^{(1)} (k_1 R_1)
- {H_m^{(1)}}' (k_1 R_2) H_m^{(2)} (k_1 R_1)
\re]
\nn  \\
&  -& n_1^2 J_m (k_2 R_2) H_m^{(1)} (k_0 R_1) \times \nn \\
& & \times \li[
{H_m^{(2)}}' (k_1 R_2) {H_m^{(1)}}' (k_1 R_1)
- {H_m^{(1)}}' (k_1 R_2) {H_m^{(2)}}' (k_1 R_1)
\re]  \nn \\
&-& n_2 J'_m (k_2 R_2) {H_m^{(1)}}' (k_0 R_1)  \times \nn \\
& & \times \li[
H_m^{(2)} (k_1 R_2) H_m^{(1)} (k_1 R_1)
- H_m^{(1)} (k_1 R_2) H_m^{(2)} (k_1 R_1)
\re] \nn \\
&+& n_1 n_2 J'_m (k_2 R_2) H_m^{(1)} (k_0 R_1)  \times \nn \\
& & \times \li[
H_m^{(2)} (k_1 R_2) {H_m^{(1)}}' (k_1 R_1)
- H_m^{(1)} (k_1 R_2) {H_m^{(2)}}' (k_1 R_1)
\re]
\nn 
\eea
for TM polarized light.
Note that Eq.~(\ref{annbtmexact}) reduces
to Eq.~(\ref{tmstartderiv}) for $n_1 = n_2$ 
when the annular billiard is reduced to a disk.

\begin{figure}[!t]
  \epsfxsize=9cm
  \centerline{\epsffile{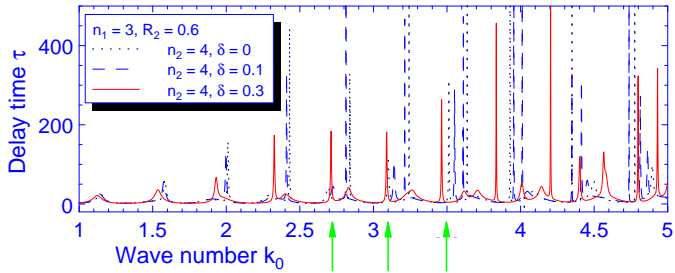}}
  \begin{center}
  \caption{Resonance peaks for increasing displacement $\delta$ in
  comparison with the concentric case (dotted) 
  for $n_2 > n_1$.
  Here, whispering gallery modes of the outer disk are 
  shifted to the left due to an increase of the effective
  refractive index. However, other modes are affected in a different way,
  as for example the 
  resonances marked by arrows.
  }
  \label{fig_delay134}
  \end{center}
\end{figure}

In order to investigate the eccentric case, we focus on the
$S$-matrix method. The derivation of the $S$-matrix for the eccentric
annular billiard is outlined in the appendix. 
As discussed in Sec.~\ref{sec_disk} the information on resonance position and
width is contained both in the complex wave vector that solves the resonance
equation deduced from Maxwell's equations and in the delay-time plot $\tau(k)$.
This is illustrated in the inset of Fig.~\ref{fig_delayconc}(a)
where the resonance positions
and widths found from $\tau(k)$ are compared with the numerically exact solutions of
Eq.~(\ref{annbtmexact})
for concentric geometries. The delay-time plot in Fig.~\ref{fig_delayconc}(a)
reveals a systematic deviation of the first few resonance positions to the
right (left), if the refractive index of the inner disk is lower
(higher) than that in the annulus. However, the deviation from the
concentric case is rather small.
It suggests that the low-lying resonances in
the (concentric) annular geometry are very similar
to the WGMs of the dielectric disk and mainly localized at the
outer boundary. 
However, the resonant
wave function {\em does} experience the change of the refractive
index in the inner disk as indicated by the shift of the resonance
position. The direction of the shift is most easily seen when
thinking in terms of an {\em effective} refractive index $n_{\rm eff}$,
\be\label{defneff}
n_{\rm eff} \df \li( 1 - \fr{R_2^2}{R_1^2} \re) n_1 + \fr{R_2^2}{R_1^2} n_2
\:.
\ee
An inner disk of lower refractive implies $n_{\rm eff} < n_1$ and
a larger spacing between the resonances. This is easily understood when
considering an eigenvalue $n k =$const. of the (closed) dielectric disk.
Obtaining the same constant value for a smaller
$n$ requires a higher $k$. In contrast, an inner disk of higher
refractive index reduces the spacing between the resonances.
This effect is strongest for resonances of high radial
quantum number $\rho$ and small angular
momentum quantum number $m$, since they rather extend to the inner
regions of the disk or the annular billiard.
(In terms of the
ray picture, they correspond to smaller angles of incidence,
leading to the same conclusion.) 
Accordingly, the
effect reduces for increasing $m$ and eventually vanishes if the
inner disk is not seen any more \cite{remark_wgm}.
In Fig.~\ref{fig_delayconc}(a) resonances are marked by arrows that exist only if
the refractive index of the inner disk is highest. One corresponding wave
pattern, together with a ray analogue, is shown in 
Fig.~\ref{fig_delayconc}(b). 
It reveals that the ``double WGM'' structure results from a star-like trajectory.

In Fig.~\ref{fig_delay134} we consider the same 
refractive indices (i.e. $n_1=3, n_2=4$) and shift now the inner disk 
off the centre. The ``double WGMs'' (again marked by arrows) are affected in
a way different from the ``conventional'' WGMs.
First of all, the systematic shift of the latter can again be
understood in terms of the
effective refractive index. The impact of an off-centered
(inner) disk  is enhanced because in the constricted region it acts like a
concentric disk with larger radius $R_2^{\rm eff} > R_2$.
Note that the resonances marked by arrows in Fig.~\ref{fig_delay134} change 
their character from ``double WGMs'' in the concentric and slightly eccentric  
cases to ``generalized'' WGMs similar to the one shown in the right panel of 
Fig.~\ref{fig_twowgms} if the symmetry breaking caused by the off-centered 
inner disk becomes too strong.  



\section{Ray-wave-correspondence for the annular billiard}
\label{sec_annbraywave}

In the previous sections we already referred to the ray-wave correspondence in
optical systems and gave several examples which were mainly based on WG modes,
and the concentric annular billiard. In this section, we first continue with
WGMs and show how they can be specifically influenced by choosing appropriate
materials. However, ray-wave correspondence holds for far more interesting
trajectories, and we will give illustrative examples how closed-billiard
trajectories are recovered in the {\em open} system using real and phase space
portraits.

\subsection{Classes of whispering gallery modes in annular systems}

\begin{figure}
  \epsfxsize=9cm
  \centerline{\epsffile{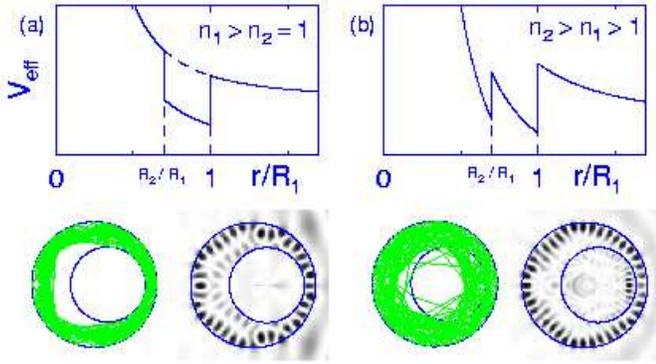}}
  \begin{center}
  \caption{
  Effective potential for the annular billiard for two different sets of
  refractive indices ($n_0$=1). In the lower panels corresponding examples of ray
  trajectories (left)
  and wave functions (right) are shown.
  The similarity of these resonances to
  whispering gallery modes suggest their classification as ``generalized'' WGMs.}
  \label{fig_twowgms}
  \end{center}
\end{figure}

In Sec.~\ref{sec_disk} we introduced the concept of the effective potential
as a wave-picture method when we established the analogy between Helmholtz and
Schr\"odinger equations. The generalization of this concept to the annular billiard
is straightforward and (in the concentric case) essentially given by the superposition
of two disks, see Eq.~(\ref{veff}). 
The result is
schematically shown in Fig.~\ref{fig_twowgms}. Again, we have to distinguish
two cases: In Fig.~\ref{fig_twowgms}(a), the refractive index $n_1$ in the annulus is
highest, whereas in Fig.~\ref{fig_twowgms}(b) the optical density increases
towards the inner disk. Consequently, in the first case, Fig.~\ref{fig_twowgms}(a),
the potential well coincides with the annular region: Rays in between the two
disks can be totally reflected at either boundary as illustrated in the lower
panels.

The situation is different when the refractive index $n_2$ of the inner disk
is highest [Fig.~\ref{fig_twowgms}(b)]. The well extends now beyond the
inner boundary to values $r<R_2$ indicating that the inner disk may support
annular WGMs in the constricted region, see the lower panels.
This is consistent with the ray picture interpretation stating that
each ray in the annulus
that hits the inner boundary will enter the inner disk. However, because of the
double-well structure of the effective potential this case is even richer:
There are modes that mainly live in one of the two wells, corresponding to WGMs
of the inner and outer disk, respectively. The height of the separating barriers depends
on the wave number, the quantum number $m$, the geometry, and in particular on
the ratio of the refractive indices that can be used to tune the height of the
barrier (note that at the same time the depths of the minima are changed).

\subsection{Towards closed systems}

\begin{figure}[!t]
  \epsfxsize=9cm
  \centerline{\epsffile{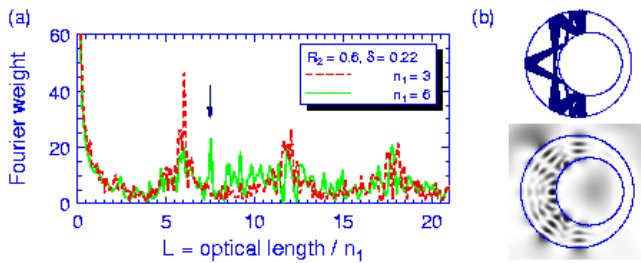}}
  \begin{center}
  \caption{
  (a) Fourier transform of the delay time $\tau(k_0)$ for the annular billiard geometry
  $R_2=0.6, \delta=0.22$ and two refractive index combinations, $n_0=n_2=1$,
  $n_1=3$ (dashed line), and $n_1=6$ (full line). The appearance of a new peak
  at larger geometric length is clearly visible. A suitable (quasi-) periodic
  orbit candidate together with a resonant state is shown in (b).
  }
  \label{fig_towardsclosed}
  \end{center}
\end{figure}

Varying the refractive index of an optical system allows one to 
describe the transition between closed and open, optical,
systems as mentioned earlier. To illustrate this fact we increase the
refractive index of the annular region that we assume to be embedded
in vacuum ($n_0=n_2=1$, wave vector $k_0=k_2$).
The length spectrum, or Fourier transform, of the delay time $\tau(k_0)$ is
shown in Fig.~\ref{fig_towardsclosed}(a) for refractive indices $n_1=3$ and $n_1=6$
(dashed and full line, respectively).
The Fourier analysis is performed in the spirit of trace formulas that provide a 
semiclassical interpretation of quantum-mechanical results in terms of
classical periodic orbits for quantum billiards. The quantitative
extension of this approach
to optical systems will require further discussion. 
Here, we are only interested in a qualitative interpretation.

We have divided the {\em optical} length
that results from the Fourier transformation by $n_1$ in order to compare both
spectra in terms of {\em geometrical} lengths $L$. The peaks in both spectra are rather
broad and correspond roughly to the
circumference of the bigger disk (and higher harmonics)
which indeed is a
typical trajectory length 
in this geometry not only for WGMs, but also for the trajectory examples shown
Figs.~\ref{fig_sosannbhw} and \ref{fig_vierer}
(trajectory parts in the inner disk
will contribute a length that has to be corrected by a factor $n_2/n_1$).
However, the 
length spectrum for $n_1=6$ shows an
additional peak (marked by the arrow) at higher $L$. 
A ray
trajectory of suitable length is shown in  Fig.~\ref{fig_towardsclosed}(b).
The Poincar\'e fingerprint of this orbit, see Fig.~\ref{fig_sosannbhw},
possesses regular islands at $\sin \chi =0$, where in the simplest interpretation
refractive escape will occur, independent on the refractive index! That modes
of this type are found for sufficiently large $n_1$, indicates that we have to
refine our interpretation. For example, we can discuss the Fresnel reflection
coefficient $R_{\perp}$ at normal incidence, $R_{\perp} = (n_0 - n_1)^2/ (n_0 + n_1)^2$
that increases as $n_1$ is increased, reaching the value one in the limit $n_1
\to \infty$, in accordance with the picture of complete internal reflection.
This explains the observed behaviour and we present more examples of
``sophisticated''
ray-wave, or classical-quantum,
correspondence in the next paragraph.

\subsection{Correspondence in real and phase space}
\label{subsec_corrrealphase}

\begin{figure}[!t]
  \epsfxsize=9cm
  \centerline{\epsffile{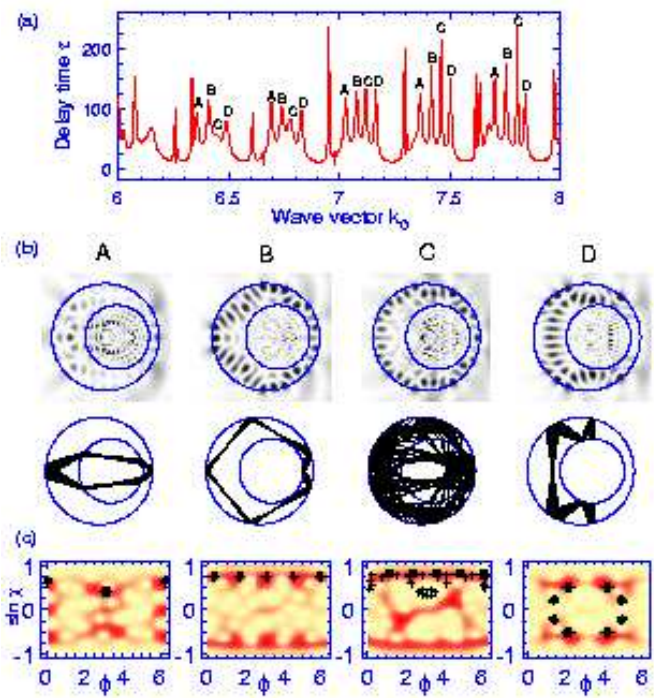}}
  \begin{center}
  \caption{
  (a) Delay time $\tau(k_0)$ for the annular billiard geometry as in Figs.~\ref{fig_sosannbhw}
  and \ref{fig_towardsclosed} 
  and $n_1=3, n_2=6$. The structure is dominated by
  groups of four resonances labelled A, B, C, D.
  The resonance at $k_0\approx 6.251$ is the one
  shown in Fig.~\ref{fig_annbgeom}(b). The four resonances of the second group
  ($6.6 < k_0 < 6.9$) are shown in (b) together with a ray trajectory
  representative. In part (c), rays (crosses) and waves 
  (intensity plot, high intensity in dark regions) are compared in phase space in
  terms of their Poincar\'e 
  and Husimi representation, respectively.
  }
  \label{fig_vierer}
  \end{center}
\end{figure}

In Fig.~\ref{fig_vierer}(a) we show a typical delay-time plot $\tau(k_0)$ for the
annular billiard with the same geometry as before 
and refractive indices $n_1=3, n_2=6$. We investigate low-lying resonances that
show a characteristic grouping of four resonances (marked by A, B, C, D) over
several periods. The corresponding wave patterns, together with suggestions for ray
analogues, are shown in Fig.~\ref{fig_vierer}(b) 
for each resonance. We have mainly chosen regular orbits as candidates because of
the regular structure of the delay-time plot.
Neighbouring resonances
of the same kind (i.e., the same letter) indeed differ by one in the number of
nodes\cite{remark_islandsize}. Note that the ray representatives stem from both
the hard-wall {\em and} the refractive billiard simulation, see Figs.~\ref{fig_sosannbhw}
and \ref{fig_refr136}.

In Fig.~\ref{fig_vierer}(c) we computed the
Husimi function \cite{doronfrischat,hackenbroich,husimi}  
for each of the wave functions \cite{romanundwir},  
and marked the rays by crosses in the
corresponding Poincar\'e SOS such that we can directly compare
the phase space presentations of waves and rays. The coincidence between regular
islands and high-probability regions (dark) of the Husimi function appears
satisfying on first sight. However, closer inspection reveals differences in
the details. For example, Husimi ``islands'' are shifted away from regular
islands as in the case of 
resonance D, with the corresponding real space modifications
[Fig.~\ref{fig_vierer}(b)] clearly visible 
as well. One possible explanation might be
provided by the Goos-H\"anchen effect that causes a lateral
shift of the reflected ray for angles of incidence around and greater than the
critical angle \cite{fresnelpre,ghslit}, thereby effectively changing
the angle of incidence.
Furthermore, we point out that the ray trajectory for resonance D is known from the hard-wall
system. The qualitative similarity to the corresponding resonant wave pattern 
is remarkable, and 
one might think of the differences as in order to meet the new interference
requirements caused by
the optical opening of the inner disk.
This gives yet another example of the predictive power of the simple ray
model when only the {\em qualitative} character of the resonances is
of interest and importance. On the other hand, it proves to be essential to
consult wave methods when one is interested in details. 



\section{Conclusions}
\label{sec_concl}

To conclude, we have investigated the ray and wave properties of
composite optical systems by applying methods known from the classical and
quantum theory of mixed dynamical systems. Using the optical annular billiard as
an example, we have shown this concept to be very fruitful. This means in
particular that already the simple ray model provides a good qualitative
understanding of the system properties, even for small wave numbers below
$nkR \approx$ 30. However, care must be taken when quantitative results are
required, or the classical (ray) phase space is directly translated into expected
wave patterns: We find regular orbits associated with regular islands in phase
space to be the dominant class of
resonant wave patterns, and suppression 
of wave functions hosted by the
chaotic part of the phase space. The dependence of this behaviour on the size
of the wave number (i.e. 
$1/\hbar$) remains an interesting
topic for future work.

One remark is due concerning the refractive indices employed in the
calculations. The index $n=3$ often used here is higher than that of
water (1.33) or glass (around 1.5 up to 1.8) but is easily reached in
semiconductor compounds where typically $n=3.3$.
An index $n=6$ seems to be presently out of reach, which, however, does not affect the
conclusions drawn here.

Summerizing, ray picture results may serve as a guide in the investigation of
wave properties of optical systems, even away from the ray limit $k \to
\infty$. For the annular billiard as an example of a compound cavity system
we demonstrated that the dominant resonant wave patterns can be seen as originating
from the regular orbits of both the 
hard-wall and the refractive billiard. 
This knowledge can be more generally used, e.g., in the construction of
microlasers
with designed properties. Knowing the potential reflection points and high-intensity
regions of modes from simple ray-based considerations allows
one to design microcavities with custemized properties. Predictions
can be made concerning,
e.g., the effective coupling between and into cavities or how to efficiently pump
lasing systems. In turn, one can think of cavity shapes designed according to the technical
requirements. The application of the ray-wave correspondence in sophisticated optical
(compound) systems therefore may provide a powerful tool for future optical
communication technologies.

Optical cavities represent interesting model systems for quantum-chaos motivated
studies. We successfully applied the $S$-matrix approach to gain spectral
information, and qualitatively discussed its periodic-orbit
interpretation \cite{dissmh}. 
The development of quantitative semiclassical theories in the spirit of the Weyl and the
trace formulas remains an open subject, in particular
for {\em compound} systems consisting of more than one region with fixed refractive
index like the annular billiard.

\begin{acknowledgments}
We thank J.~U.~N\"ockel for an introduction into the subject of optical
cavities and
acknowledge many useful discussions with T.~Dittrich, S.~Fishman, G.~Hackenbroich,
J.~U.~N\"ockel, H.~Schomerus, H.~Schanz,
R.~Schubert, P.~Schlagheck, U.~Smilansky, and J.~Wiersig. M.~H. thanks
U.~Smilansky for his hospitality at the  Weizmann Institute.
\end{acknowledgments}

\begin{appendix}
\section{$S$-matrix for the annular billiard
\label{sec_app}}
We will generalize the ideas developed in
Sec.~\ref{subsec_smatdisk} to the dielectric annular billiard in
order to determine the $S$-matrix for the {\em eccentric} annular
billiard.
This problem can be
divided into the scattering problem at the {\em outer} boundary
(between refractive indices $n_0$ and $n_1$)
and that at the {\em inner} boundary
(between indices $n_1$ and $n_2$). Although the
scattering at a dielectric disk was solved in
Sec.~\ref{subsec_smatdisk}, the situation we are confronted with here is
more complicated: the two disks lay one in the
other, and their centres will in general not coincide. 

We will begin with the 
scattering problem at the inner boundary and express
the $S$-matrix $S^{i}$ of the dielectric disk
with respect to a coordinate system with origin
displaced from the 
centre of the inner disk. This implies that
$S^{i}$ is not diagonal.
From Sec.~\ref{subsec_smatdisk} we already know 
the (diagonal) $S$-matrix $S^{ic}$ of the inner disk in {\em primed} coordinates
(see Fig.~\ref{fig_annbgeom}), Eq.~(\ref{scattmatdisk}).
We will now derive the relation between $S^{ic}$
and $S^{i}$.

\begin{figure}[!t]
  \epsfxsize=6cm
  \centerline{\epsffile{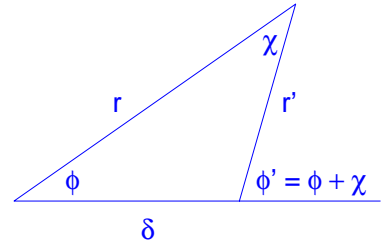}}
  \begin{center}
  \caption{ Addition theorem for Bessel functions.}
  \label{fig_addbessel}
  \end{center}
\end{figure}

To this end we write the ansatz for the wave function
in the annulus in primed coordinates, $\vec{r'} =
\vec{r}-\vec{\delta}$, with $\vec{\delta}$ being the vector from the
centre of the large disk to the centre of the smaller disk,
as
\bea
\Psi^{1c} (\vec{r}-\vec{\delta}) & = & \sum_{l= -\infty}^{\infty} a^c_l
                \li[
                H_l^{(2)} (k_1 |\vec{r}-\vec{\delta} |)
                e^{i l \phi} \re. \nn \\
                && \li. +
                \sum_{l'= -\infty}^{\infty} S^{ic}_{l l}
                H_{l'}^{(2)} (k_1 |\vec{r} -\vec{\delta} |)
                e^{i l' \phi}
                \re] \:, \label{Psianncent}
\eea
where the coefficients $a^c_l$ are to be chosen to yield the desired kind of incident wave.
We use
the addition theorem for Bessel functions
$Z_m \in \{J_m, Y_m, H_m^{(1)}, H_m^{(2)} \}$ \cite{gradstein} to relate
the arguments $ w r$ to $w r'$ ($w$ is a constant factor,
and we assume$R_2 > \delta$),
see Fig.~\ref{fig_addbessel}, 
\be\label{addthbessel}
Z_m (w r') e^{i m \phi} =
\sum_{k= -\infty}^{\infty}
J_k(w \delta) Z_{m+k}(w r) e^{i (m+k) \phi} \:.
\ee
Inserting this into Eq.~(\ref{Psianncent}), we obtain the
expression
\bea
\widetilde{\Psi}^1 &=& \sum_{l= -\infty}^{\infty}
                 \sum_{k= -\infty}^{\infty} a^c_l
                \li[
                H_{l+k}^{(2)}(k_1 r) +
                S^{ic}_{ll} H_{l+k}^{(1)}(k_1 r)
                \re]\times \nn \\
                && \times
                J_k(k_1 \delta) \: e^{i (l+k) \phi} 
\eea
for the wave function in the annulus, now expressed with respect to the
centre of the larger disk, i.e., in unprimed coordinates.
We specify the coefficients 
$a^c_l$ 
by the requirement that the amplitude
of an incident wave with angular momentum $m$ shall be normalized to one
in {\em unprimed} coordinates, 
\[
\sum_{l= -\infty}^{\infty}
                 \sum_{k= -\infty}^{\infty} a^c_l
                H_{l+k}^{(2)}(k_1 r)
                J_k(k_1 \delta) \: e^{i (l+k) \phi}
\equiv
H_m^{(2)}(k_1 r) \: e^{i m \phi} \:.
\]
With $\mu \equiv l+k$, 
and $\sum_k J_{m-(\mu-k)} J_k = \delta_{m \mu}$
we find that choosing
\be
{}^ma^c_l = J_{m-l} (k_1 \delta) \quad \forall \:\, l
\ee
provides a suitable set of coefficients for a given $m$.
Accordingly, we write
\bea
\widetilde{\Psi}^1
& =&\sum_{m= -\infty}^{\infty}
    \Bigg\{
         \sum_{\mu, k = -\infty}^{\infty}
         \li[
         \delta_{m \mu}
         H_{\mu}^{(2)}(k_1 r) e^{i \mu \phi}
         \re.  \nn  \\
& &         \li.
         +
         J_{m-(\mu -k)}(k_1 \delta) S^{ic}_{(\mu-k)(\mu-k)} J_k(k_1 \delta)
          H_{\mu}^{(1)}(k_1 r) e^{i \mu \phi}
                \re]
                \Bigg\} \nn \\
& \equiv & \sum_{m= -\infty}^{\infty}
        \li\{
        H_{m}^{(2)}(k_1 r) e^{i m \phi}
        +
        \sum_{\mu= -\infty}^{\infty}
        S^{i}_{m \mu}  H_{\mu}^{(1)}(k_1 r) e^{i \mu \phi}
        \re\}  \nn
\eea
where we have read off the scattering matrix $S^{i}$ of the inner
disk with respect to the centre of the outer disk,
\be\label{s2unprimed}
S^{i}_{m \mu} \df
\sum_{k=\infty}^{\infty}
J_{m-(\mu -k)}(k_1 \delta) S^{ic}_{(\mu-k)(\mu-k)} J_k(k_1 \delta)
\:.
\ee
The structure of this equation suggests a notation in terms of
a transformation matrix $U$, namely $S^{i} = U^{-1} S^{ic} U$,
that describes the change in the origin of the coordinate
system. 
We find $U_{l' l} = J_{l'-l}$ and
$U^{-1}_{l' l} = J_{l-l'}$.

The scattering matrix $S^{i}$ allows us to describe
the scattering at an off-centred disk,
and we can now formulate the scattering problem of the annular
billiard in the spirit of Sec.~\ref{subsec_smatdisk}.
Accordingly, we start with an ansatz
for the wave function $\Psi^0$ outside the
annular system ($|\vec{r}| > R_1$, using polar coordinates)
of the form
\bea
\Psi^0 (\vec{r}) &=& \sum_{M=-\infty}^{\infty} \Psi^0_M (\vec{r}) \nn \\
        & = &  \sum_{M=-\infty}^{\infty}
                \li [
                \Psi_M^- (k_0 \vec{r}) +
                \sum_{M'= -\infty}^{\infty} S_{MM'} \Psi_{M'}^+(k_0 \vec{r})
                \re] \label{Psiout}\:, \nn
\eea
where we have introduced the scattering matrix $S$ of the
(compound) system and the definitions 
\bea
\Psi_M^- (k_0 \vec{r}) & = & H_M^{(2)}(k_0 r) e^{i M \phi } \:,
\label{psimminus} \\
\Psi_M^+ (k_0 \vec{r}) & = & H_M^{(1)}(k_0 r) e^{i M \phi } \:,
\label{psimplus}
\eea
for incoming and outgoing waves outside the disk. Note that we
have used 
the freedom in fixing one of the amplitudes.

Similarly, we write for the wave function $\Psi^1$ in the annular
region
\be\label{Psiann}
\Psi^1 (\vec{r}) = \sum_{l= -\infty}^{\infty} a_l
                \li[
                \Psi_l^- (k_1 \vec{r}) +
                \sum_{l'= -\infty}^{\infty} S^{i}_{ll'} \Psi_{l'}^+(k_1 \vec{r})
                \re] \:,
\ee
with the amplitudes $a_l$, the abbreviations as in
Eqs.~(\ref{psimminus}, \ref{psimplus}), and $S^{i}$ from Eq.~(\ref{s2unprimed}).

Now, we determine $S$ from the matching conditions,
introduce the notation of capital letters
for functions of argument $k_0 r$, and reserve lower case characters for
the argument $k_1 r$. Given an incident wave of angular momentum $M$,
wave function matching for each
angular momentum $L$ of the scattered waves yields
\bea
& & H_M^{(2)} e^{i M \phi} \delta_{ML} + S_{ML} H_L^{(1)} e^{i L \phi}
 \nn \\
& & \quad \quad \quad =  a_L^{(M)} h_L^{(2)} e^{i L \phi} +
\sum_{l= -\infty}^{\infty} a_l^{(M)} S^{i}_{l L }h_L^{(1)} e^{i L \phi}
\:, \nn
\eea
where 
the amplitudes $a_j^{(M)}$ are 
coefficients associated with an incoming function of angular
momentum $M$, namely $H_M^{(2)}$. Since this has to hold for all
$M$, and at fixed $M$ for all $L$, we write this as a matrix
equation
\be\label{matchannbfct}
\la {}^{(M)} H^{(2)} | + \la S^{(M)} | H^{(1)} =
\la a^{(M)} | \li( h^{(2)} + S^{i} h^{(1)} \re) \:,
\ee
where $S^{i}$ is a matrix, $h^{(2)}$ and $h^{(1)}$ are diagonal
matrices, $h^{(1,2)}_{lj} = h^{(1,2)}_l \delta_{lj}$,
and we adopt the {\em bra}-notation for quantities
that, at fixed $M$, are
transposed vectors and gain matrix character once
$M$ is varied.
With this notation we immediately write the matching condition for the
derivatives as
\bea
& & k_0 \li( \la {}^{(M)} {H}^{(2)'} | + \la S^{(M)} | {H}^{(1)'} \re) \
\nn \\
& & \quad \quad \quad =
\la a^{(M)} | k_1 \li( {h^{(2)}}' + S^{i} {h^{(1)}}' \re) \:.
\nn 
\eea
From Eq.~(\ref{matchannbfct}) we find after substituting
$F \equiv h^{(2)} + S^{i} h^{(1)} $  that
\[
\la a^{(M)} | = \li( \la {}^{(M)} H^{(2)} | + \la S^{(M)} | H^{(1)} \re) \, F^{-1}
\:.
\]
Introducing furthermore
$F' = {h'}^{(2)} + S^{i} {h'}^{(1)} $ and $W = F^{-1} F'$, we
write the $S$-matrix solution of the problem as
\[
S =  \li( k_1 \, H^{(2)} \, W - k_0 \, {H'}^{(2)} \re)
     \li( k_0 \, {H'}^{(1)} - k_1 \, H^{(1)} \, W \re)^{-1} \:.
\]
This last equation 
allows us to apply the Wigner-delay-time
approach to resonances, cf.~Sec.~\ref{subsec_smatdisk}, and we used this method
to study resonances of the optical annular billiard,
cf.~Secs.~\ref{sec_annbpictures} and \ref{sec_annbraywave}.

We complete the discussion here with 
some comments on the wave functions.
We have not yet given the wave function in the
inner disk. The ansatz is 
a sum over Bessel functions,
\[
\Psi^{2c} (\vec{r'}) = 
\sum_{l= -\infty}^{\infty} b^c_l J_l(k_2 r') e^{i l \phi'} \:,
\]
where we adopted primed coordinates for convenience. The
coefficients $b_l$ are found from matching with the wave function
in the annulus at the {\em inner} boundary. To this end we have
to rewrite the annular wave function (\ref{Psiann}) in terms of primed
coordinates by applying the addition theorem (\ref{addthbessel}) for
Bessel functions. After straightforward algebra we find
\[
\Psi^{1c}(\vec{r'}) =
\sum_{l= -\infty}^{\infty} a^c_l
\li[ H_l^{(2)}(k_1 r') + S^{ic}_{ll}  H_l^{(1)}(k_1 r') \re]
e^{i l \phi'} \:,
\]
where the coefficients $a_l^c$ are related to the $a_l$ by
$a_l^c = \sum_{l'= -\infty}^{\infty} a_{l'} J_{l'-l}$. 

Another remark is in order concerning the validity of the
addition-theorem-based expansion of the Bessel function when changing
between primed and unprimed coordinates. 
Expansion of
the annular wave function in {\em primed} coordinates fails near
the outer boundary where $|\vec{r'}| > R_1 - R_2 -\delta$.
Similarly, expanding the annular wave function
in {\em unprimed} coordinates does not work
near the inner boundary where $|\vec{r}| < R_2 + \delta$.
The reason for this behaviour is that angular momentum is not
conserved in the eccentric annular billiard, and the expansion
breaks down at radii where waves explore this symmetry-breaking region
because the corresponding interface boundaries are hit.

\end{appendix}




\begin{references}
\bibitem{remark_doublet} For such open annular billiards
questions concerning the doublet-splitting addressed
in \cite{bohigas,doronfrischat} are not of relevance
because the splitting it masked by resonance broadening.
\bibitem{remark_polar}
The vector character of light requires to distinguish between two
polarizations. In our planar systems, either the transverse magnetic
(TM) or electric (TE) field can lie in the system plane.
\bibitem{remark_radschroed}
The first derivative 
in the first term can be eliminated by a 
substitution $\xsi = \ln (kr)$.
\bibitem{remark_resolaser}
For microlaser operation, resonances of intermediate width are of particular
interest: For too broad resonances leakage ceases the crossing of the lasing
threshold, whereas very narrow resonances will fail to provide sufficient output
power.
\bibitem{remark_geomparam} The geometry parameters mainly studied here were
chosen because of the particularly interesting phase space structure.
Generalization of the results to other geometries is straightforward.
\end{references}
\end{document}